\documentstyle[preprint,aps,pre]{revtex}
\begin{document}
\title{Shear viscosity for a heated granular binary mixture at low-density}
\author{Jos\'e Mar\'{\i}a Montanero\footnote[1]
{Electronic address: jmm@unex.es}}
\address{Departamento de Electr\'onica e Ingenier\'{\i}a Electromec\'anica,
Universidad de Extremadura, E-06071 Badajoz, Spain}
\author{Vicente Garz\'{o}\footnote[2]{Electronic address: vicenteg@unex.es}}
\address{Departamento de F\'{\i}sica, Universidad de Extremadura, E-06071
Badajoz, Spain}

\date{\today}
\maketitle

\begin{abstract} 

The shear viscosity for a heated granular binary mixture of smooth hard 
spheres at low-density is analyzed. The mixture is heated by the action of 
an external driving force (Gaussian thermostat) which exactly compensate for 
cooling effects associated with the dissipation of collisions. The study is 
made from the Boltzmann kinetic theory, which is solved by using two 
complementary approaches. First, a normal solution of the Boltzmann 
equation via the Chapman-Enskog method is obtained up to first order in 
the spatial gradients. The mass, heat, and momentum fluxes are determined 
and the corresponding transport coefficients identified. As in the free 
cooling case [V. Garz\'o and J. W. Dufty, Phys. Fluids {\bf 14}, 1476 
(2002)], practical evaluation requires a Sonine polynomial approximation, 
and here it is mainly illustrated in the case of the shear viscosity. Second, to check 
the accuracy of the Chapman-Enskog results, the 
Boltzmann equation is numerically solved by means of the Direct Simulation 
Monte Carlo (DSMC) method. The simulation is performed for a system under 
uniform shear flow, using the Gaussian thermostat to control inelastic 
cooling. The comparison shows an excellent agreement between theory and 
simulation over a wide range of values of the restitution coefficients and 
the parameters of the mixture (masses, concentrations, and sizes). 
 
\end{abstract} 
 
\draft 
\pacs{PACS number(s): 45.70.Mg, 05.20.Dd, 51.10.+y} 
 
\bigskip \narrowtext 
 
\section{Introduction}
\label{sec1}

The macroscopic behavior of rapid granular flows can be described  
through hydrodynamic equations accounting for dissipation among the 
interacting particles. A basis for the derivation of the hydrodynamic 
equations and explicit expressions for the transport coefficients appearing 
in them is provided by the corresponding Boltzmann kinetic theory in 
the low-density regime. In the simplest model the grains are taken to be 
smooth hard spheres with inelastic collisions. Assuming the existence of a 
normal solution for sufficiently long space and time scales, the 
Chapman-Enskog method\cite{CC70} can be applied to get the velocity distribution 
function in terms of the hydrodynamic fields and their spatial gradients. 
This method must be conveniently adapted to inelastic collisions due to the 
new time dependence of temperature resulting from collisional energy loss. 
In the case of a monocomponent gas, the Navier-Stokes transport coefficients 
have been obtained in terms of the restitution coefficient without 
limitation on the degree of inelasticity\cite{BDKS98,BC01,GM02}. This 
analysis for a monocomponent system has been also extended to dense gases 
in the context of the Enskog equation\cite{GD99a}.

Similar studies for multicomponent granular gases are more scarce and 
most of them limited to the asymptotically weak dissipation 
limit\cite{varios}. Although these studies permit in 
principle different temperatures for each species, they usually assume energy 
equipartition and so the partial temperatures $T_i$ are made equal to the mixture 
temperature $T$. However, some recent results obtained in molecular dynamics
simulations \cite{DHGD02} as well as in real experiments of vibrated 
mixtures in three \cite{WP02} and two \cite{FM02} dimensions clearly show 
the breakdown of energy equipartition. A more recent kinetic 
theory calculation which takes into account temperature differences has been 
carried out by Garz\'o and Dufty \cite{GD02}. They solved the 
set of Boltzmann coupled equations of the binary mixture by means of the 
Chapman-Enskog expansion for states near the local homogeneous cooling 
state. The mass, heat, and momentum fluxes were determined to first 
order in the gradients of the hydrodynamic fields and the associated 
transport coefficients were explicitly identified. As in the case of elastic 
collisions\cite{CC70}, these transport coefficients verify a set of 
coupled linear integral equations which are solved approximately by using 
the leading terms in a Sonine polynomial expansion. The results derived 
by Garz\'o and Dufty \cite{GD02} provide a description 
of hydrodynamics in binary granular mixtures valid a priori  
over the broadest parameter range and not limited to the quasielastic regime. In 
particular, the consequences of the temperature differences on the transport 
coefficients were shown to be quite significant.

In the case of molecular fluid mixtures, it is known that the leading order 
truncation is quite accurate, except for extreme mass ratios (e.g., 
electron--proton systems). Much less is known in the case of inelastic 
collisions, although some comparisons with computer simulations for 
homogeneous states indicate that the accuracy is similar to that 
for elastic collisions\cite{MG02,BT02}. The objective here is to compare the 
kinetic theory predictions for the shear viscosity with those obtained from 
a numerical solution of the Boltzmann equation by means of the 
Direct Simulation Monte Carlo (DSMC) method\cite{B94}. Specifically, the 
simulations are performed for a granular mixture undergoing uniform shear 
flow (USF), namely, a macroscopic state characterized by constant 
partial densities $n_i$ ($i=1,2$), uniform temperature $T$ and a linear 
flow velocity profile $u_{i,x}=u_x=ay$, $a$ being the constant shear rate. 
In a molecular fluid under USF, the temperature increases in time due to 
viscous heating. As a consequence, the average collision frequency $\nu$ 
(which is proportional to $T^{1/2}$ for hard 
spheres) increases with time and the reduced shear rate $a^*=a/\nu$ 
(which is the relevant nonequilibrium parameter of the problem) tends to 
zero in the long time limit. This implies that for sufficiently long times 
the system reaches a regime described by linear hydrodynamics and the 
Navier-Stokes shear viscosity can be identified \cite{NO79,MS96}.  For granular 
fluids, the inelasticity of collisions introduces an energy sink in the 
balance equation for the temperature. Thus, the relationship between the 
temperature and the shear viscosity is not as simple as for molecular fluids 
since there is a competition between viscous heating and collisional 
cooling. However, if the granular fluid is externally excited by an external 
energy source that exactly compensates for the collisional energy loss, the 
viscous heating effect is still able to heat the system (as in the elastic 
case) and one can identify the shear viscosity in the limit of small shear rate 
relative to the collision frequency $\nu$ (i.e., $a^*\to 0$). 
Although there are several choices for the external 
driving force, here we consider an external thermostat proportional to the 
peculiar velocity. This thermostat has been frequently used in 
nonequilibrium molecular dynamics simulations of molecular fluids\cite{EM90} 
and has the advantage that, in the absence of shear, it does not affect the 
dynamics of the system at all since it is formally equivalent to a rescaling of the 
velocities\cite{DHGD02} .

The motivation of our study is twofold. First,  in light of some doubts about the 
validity of a hydrodynamic description for granular flow, the 
comparison with simulation allows us to test the Chapman-Enskog solution obtained by 
assuming the existence of a {\em normal} or hydrodynamic regime. Since the 
parameter space here is quite large the tests of the theory and concepts are 
quite stringent. Second, as said above, we can also assess 
the degree of reliability of the approximate solution (first Sonine 
polynomial approximation) to the resulting integral equation over a wide 
range of the parameter space. With respect to the driving external force used in our 
analysis, we do not claim that it is the most suited one to model any real 
experiment. However, it has the advantage that it can be incorporated into 
kinetic theory and computer simulations very easily  and it allows to check 
the assumptions of the Chapman-Enskog method.

The plan of the paper is as follows. In Section \ref{sec2}, we review the 
Boltzmann equation and associated macroscopic conservation laws in the 
presence of the Gaussian thermostat. The Chapman-Enskog method is applied in 
Section \ref{sec3} to get all the transport coefficients of the mixture, 
with special emphasis in the shear viscosity coefficient. The details of the 
derivation are displayed in Appendix \ref{appA}. Section \ref{sec4} deals 
with the application of the DSMC method of the Boltzmann equation to USF 
with thermostat. The Chapman-Enskog and simulation results are compared in 
Section \ref{sec5} at the level of the shear viscosity  showing a good agreement. 
We close the paper in Section \ref{sec6} with a discussion of the results presented.

\section{The Boltzmann equation and transport coefficients}
\label{sec2}
 
We consider a binary mixture of smooth hard spheres of 
masses $m_{1}$ and $m_{2} $, and diameters $\sigma _{1}$ and $\sigma _{2}$. The inelasticity of 
collisions among all pairs is characterized by three independent constant 
coefficients of normal restitution $\alpha_{11}$, $\alpha_{22}$, and 
$ \alpha_{12}=\alpha_{21}$, where $\alpha_{ij}$ is the restitution 
coefficient for collisions between particles of species $i$ and $j$. 
Due to the intrinsic dissipative character of collisions, an energy supply 
is requested to fluidize a granular gas. For simplicity, here the 
fluidization is driven by the action of a non-conservative external force, 
frequently referred to as the Gaussian thermostat. In this case, the 
mixture is heated by an ``anti-drag'' force, linear in the peculiar velocity 
${\bf V}$ and chosen to exactly compensate for 
collisional cooling. As said in the Introduction, this deterministic 
thermostat has been widely used in computer simulations of molecular fluids 
\cite{EM90}. Under these conditions and in the low-density regime, the 
distribution functions $f_{i}({\bf r},{\bf v};t)$ $
(i=1,2)$ for the two species are determined from the set of nonlinear 
Boltzmann equations 
\begin{equation} 
\left( \partial _{t}+{\bf v}_{1}\cdot \nabla \right) 
f_{i}+\frac{1}{2}\xi
\frac{\partial}{\partial {\bf v}_1}\cdot \left({\bf V}_1f_i\right)
=\sum_{j}J_{ij}\left[ {\bf v}_{1}|f_{i}(t),f_{j}(t)\right] \;,
\label{2.1} 
\end{equation} 
where the constant $\xi$ is taken to be the same for 
each species \cite{DHGD02,BT02,HBB00}. Here, ${\bf V}_1\equiv 
{\bf v}_1-{\bf u}$, ${\bf u}$ being the flow velocity. The Boltzmann 
collision operator $J_{ij}\left[ {\bf v}_{1}|f_{i},f_{j}\right] $ describing 
the scattering of pairs of particles is  
\begin{eqnarray} 
J_{ij}\left[ {\bf v}_{1}|f_{i},f_{j}\right] &=&\sigma _{ij}^{2}\int d{\bf v} 
_{2}\int d\widehat{\bbox {\sigma }}\,\Theta (\widehat{{\bbox {\sigma }}} 
\cdot {\bf g}_{12})(\widehat{\bbox {\sigma }}\cdot {\bf g}_{12})  \nonumber 
\\ &&\times \left[ \alpha _{ij}^{-2}f_{i}({\bf r},{\bf v}_{1}^{\prime 
},t)f_{j}( {\bf r},{\bf v}_{2}^{\prime },t)-f_{i}({\bf r},{\bf v}
_{1},t)f_{j}({\bf r}, {\bf v}_{2},t)\right] \;,  \label{2.2} 
\end{eqnarray} 
where $\sigma _{ij}=\left( \sigma _{i}+\sigma _{j}\right) /2$, $\widehat{
\bbox {\sigma}}$ is a unit vector along their line of centers, $\Theta $ is 
the Heaviside step function, and ${\bf g}_{12}={\bf v}_{1}-{\bf v}_{2}$. The 
primes on the velocities denote the initial values $\{{\bf v}_{1}^{\prime},  
{\bf v}_{2}^{\prime }\}$ that lead to $\{{\bf v}_{1},{\bf v}_{2}\}$ 
following a binary collision:  
\begin{equation} 
{\bf v}_{1}^{\prime }={\bf v}_{1}-\mu _{ji}\left( 1+\alpha _{ij}^{-1}\right) 
(\widehat{{\bbox {\sigma }}}\cdot {\bf g}_{12})\widehat{{\bbox {\sigma }}} 
,\quad {\bf v}_{2}^{\prime }={\bf v}_{2}+\mu _{ij}\left( 1+\alpha 
_{ij}^{-1}\right) (\widehat{{\bbox {\sigma }}}\cdot {\bf g}_{12})\widehat{ 
\bbox {\sigma}} , \label{2.3} 
\end{equation} 
where $\mu _{ij}=m_{i}/\left( m_{i}+m_{j}\right)$.

The relevant hydrodynamic fields are the number densities $n_{i}$, 
the flow velocity ${\bf u}$, and the ``granular'' temperature $T$. 
They are defined in terms of moments of the distributions $f_{i}$ as  
\begin{equation} 
n_{i}=\int d{\bf v}f_{i}({\bf v})\;,\quad \rho {\bf u}=\sum_{i}\int 
d {\bf v}m_{i}{\bf v}f_{i}({\bf v})\;,  \label{2.4} 
\end{equation} 
\begin{equation} 
nT=p=\sum_{i}\int d{\bf v}\frac{m_{i}}{3}V^{2}f_{i}({\bf v})\;, 
\label{2.5} 
\end{equation} 
where $n=n_{1}+n_{2}$ is the total number density, $\rho 
=m_{1}n_{1}+m_{2}n_{2}$ is the total mass density, and $p$ is the 
hydrostatic pressure. At a kinetic level, it is convenient to introduce the 
kinetic temperatures $T_i$ for each species defined as 
\begin{equation}
\label{2.6}
\frac{3}{2}n_iT_i=\int d{\bf v}\frac{m_{i}}{2}V^{2}f_{i}.
\end{equation}

The collision operators conserve the particle number of each species and the 
total momentum but the total energy is not conserved:
\begin{equation}
\label{2.7}
\int d{\bf v}J_{ij}[{\bf v}|f_i,f_j]=0,\quad 
\sum_{i,j}\int d{\bf v}m_i{\bf v}J_{ij}[{\bf v}|f_i,f_j]=0,
\end{equation}
\begin{equation}
\label{2.8}
\sum_{i,j}\int d{\bf v}\case{1}{2}m_i{\bf v}^2J_{ij}[{\bf 
v}|f_i,f_j]=-\case{3}{2}nT\zeta,
\end{equation}
where $\zeta$ is identified as the {\em cooling rate} due to inelastic 
collisions among all species. The macroscopic balance equations follow 
from the Boltzmann equation (\ref{2.1}) and Eqs.\ (\ref{2.7}) and 
(\ref{2.8}). They are given by  
\begin{equation} 
D_{t}n_{i}+n_{i}\nabla \cdot {\bf u}+\frac{\nabla \cdot {\bf j}_{i}}{m_{i}} 
=0\;,  \label{2.9} 
\end{equation} 
\begin{equation} 
D_{t}{\bf u}+\rho ^{-1}\nabla {\sf P}=0\;,  \label{2.10} 
\end{equation} 
\begin{equation} 
D_{t}T-\frac{T}{n}\sum_{i}\frac{\nabla \cdot {\bf j}_{i}}{m_{i}}+\frac{2}{3n} 
\left( \nabla \cdot {\bf q}+{\sf P}:\nabla {\bf u}\right) 
=-(\zeta-\xi)T\;. \label{2.11} 
\end{equation} 
In the above equations, $D_{t}=\partial _{t}+{\bf u}\cdot \nabla $ is the 
material derivative,  
\begin{equation} 
{\bf j}_{i}=m_{i}\int d{\bf v}\,{\bf V}\,f_{i}({\bf v})
\label{2.12} 
\end{equation} 
is the mass flux for species $i$ relative to the local flow,  
\begin{equation} 
{\sf P}=\sum_{i}\,\int d{\bf v}\,m_{i}{\bf V}{\bf V}\,f_{i}({\bf  
v})  \label{2.13} 
\end{equation} 
is the total pressure tensor, and  
\begin{equation} 
{\bf q}=\sum_{i}\,\int d{\bf v}\,\case{1}{2}m_{i}V^{2}{\bf V} 
\,f_{i}({\bf v})  \label{2.14} 
\end{equation} 
is the total heat flux.

The energy balance equation (\ref{2.11}) shows that 
the existence of a driving with the choice $\xi=\zeta$ compensates for 
the cooling effect due to the inelasticity of collisions. In that case, the 
macroscopic balance equations look like those of a conventional mixture with 
elastic collisions. Nevertheless, the transport coefficients entering in the 
constitutive equations are in general different from those of a gas of 
elastic particles.  Furthermore, for systems with elastic collisions, the specific set of 
gradients contributing to each flux is restricted by fluid symmetry, time reversal 
invariance (Onsager relations), and the form of the entropy porduction\cite{GM84}. In the 
case of inelastic collisions only fluid symmetry applies and so there is  more flexibility 
in representing the fluxes and identifying the corresponding transport coefficients. 
It follows from fluid symmetry that the pressure tensor has the same form to first order 
in the gradients as for the monocomponent gas. 
In the case of heat and mass fluxes, several different (but equivalent) choices of hydrodynamic 
fields can be used and some care is required in comparing transport coefficients in the 
different representations.  Here, as done in the unforced 
case  \cite{GD02}, we take the gradients of the mole fraction $x_1=n_1/n$, 
the pressure $p$, the temperature $T$ and the flow velocity ${\bf u}$ as 
the relevant ones.  Thus, in this representation, the phenomenological constitutive 
relations for the fluxes in the low-density regime have the forms\cite{Z74}
\begin{equation}
\label{2.15}
{\bf j}_1=-\frac{m_1m_2n}{\rho}D\nabla x_1-\frac{\rho}{p}D_p\nabla p-
\frac{\rho}{T}D'\nabla T,\quad {\bf j}_2=-{\bf j}_1,
\end{equation}
\begin{equation}
\label{2.16}
{\bf q}=-T^2D''\nabla x_1-L\nabla p-\lambda\nabla T,
\end{equation}
\begin{equation}
\label{2.17}
P_{\alpha\beta}=p\delta_{\alpha\beta}-\eta\left(\nabla_\beta u_\alpha+
\nabla_\alpha u_\beta-\frac{2}{3}\delta_{\alpha\beta}\nabla \cdot {\bf 
u}\right).
\end{equation}
The transport coefficients are the diffusion coefficient $D$, the thermal 
diffusion coefficient $D'$, the pressure diffusion coefficient $D_p$, the 
Dufour coefficient $D''$, the thermal conductivity $\lambda$, the pressure 
energy coefficient $L$, and the shear viscosity $\eta$. The Chapman-Enskog 
method \cite{CC70} generalized to inelastic collisions allows one to get 
explicit expressions for these transport coefficients as functions of the 
restitution coefficients and the parameters of the mixture.

\section{Shear viscosity of a heated granular mixture}
\label{sec3} 

The Chapman-Enskog method assumes the existence of a {\em normal} solution 
in which all space and time dependence of the distribution function occurs 
through a functional dependence on the hydrodynamic fields
\begin{equation}
\label{3.1}
f_i({\bf r}, {\bf v}_1,t)=f_i[{\bf v}_1|x_1({\bf r},t), p({\bf r},t), 
T({\bf r},t), {\bf u}({\bf r},t)].
\end{equation}  
This functional dependence can be made local in space and time by means of 
an expansion in gradients of the fields. Thus, we write $f_{i}$ as a series 
expansion in a formal parameter $\epsilon $ measuring the nonuniformity of 
the system,  
\begin{equation} 
f_{i}=f_{i}^{(0)}+\epsilon \,f_{i}^{(1)}+\epsilon^2 \,f_{i}^{(2)}+\cdots \;, 
\label{3.2} 
\end{equation} 
where each factor of $\epsilon $ means an implicit gradient of a 
hydrodynamic field. The local reference state $f_i^{(0)}$ is chosen to give 
the same first moments as the exact distribution $f_i$. The time derivatives 
of the fields are also expanded as $\partial_t=\partial_t^{(0)}+\epsilon 
\partial_t^{(1)}+\cdots$. The coefficients of the time derivative expansion 
are identified from the balance equations (\ref{2.9})--(\ref{2.11}) after 
expanding the fluxes, and the cooling rate $\zeta$ in a similar series as 
(\ref{3.2}). More details on the Chapman-Enskog method adapted to inelastic 
collisions can be found in Ref.\ \onlinecite{GD02}. Now, the main difference 
with respect to the free cooling case \cite{GD02} is that the sink term in 
the energy equation is zero (when one takes $\xi=\zeta$), so that the terms 
coming from the time derivative $\partial_t^{(0)}$ vanish.

In the zeroth order, $f_i^{(0)}$ obeys the kinetic equation 
\begin{equation}
\label{3.3}
\frac{1}{2}\zeta^{(0)}\frac{\partial}{\partial {\bf V}}
\cdot \left({\bf V}f_i^{(0)}\right)=\sum_{j}J_{ij}[f_i^{(0)},f_j^{(0)}],
\end{equation}
where use has been made of the choice $\xi^{(0)}=\zeta^{(0)}$. Here, 
$\zeta^{(0)}$ is determined by Eq.\ (\ref{2.8}) to zeroth order.  With 
this choice, Eq.\ (\ref{3.3}) is {\em identical} to the the one obtained in 
the unforced case \cite{GD99b}, and there is an exact 
correspondence between the homogeneous cooling state and this type of driven 
steady state. This is one of the advantages of the Gaussian thermostat. 
Dimensional analysis requires that $f_i^{(0)}({\bf V})$ must be of the form 
\begin{equation}
\label{3.3.1}
f_i^{(0)}(V)=n_iv_0^{-3}\Phi_i(V/v_0)
\end{equation}
where 
\begin{equation}
\label{3.3.2}
v_0=\sqrt{2T\frac{m_1+m_2}{m_1m_2}}
\end{equation}
is a thermal velocity defined in terms of the temperature $T$ of the 
mixture. So far, the exact form of $\Phi_i$ has not been found, although a 
good approximation for thermal velocities can be obtained from an expansion 
in Sonine polynomials\cite{GD99b}. In the leading order, $\Phi_i$ is given 
by 
\begin{equation}
\label{3.4}
\Phi_i(V^*)\to \left(\frac{\theta_i}{\pi}\right)^{3/2}e^{-\theta_iV^{*2}}
\left[1+\frac{c_i}{4}\left(\theta_iV^{*4}-5\theta_iV^{*2}+\frac{15}{4}\right)
\right],
\end{equation}
where $V^*=V/v_0$, $\theta_i=(\mu_{ji}\gamma_i)^{-1}$, and $\gamma_i=T_i/T$. 
The coefficients $c_i$ (which measure the deviation of $\Phi_i$ from the 
reference Maxwellian) are determined consistenly from the Boltzmann 
equation. The approximation (\ref{3.4}) provides detailed predictions for 
the cooling rate $\zeta^{(0)}$, the temperature ratio $T_1/T_2$ and the 
cumulants $c_i$ as functions of the mass ratio, size ratio, composition and 
restitution coefficients. Recently, the accuracy of this approximate 
solution has been confirmed by Monte Carlo simulation of the 
Boltzmann equation over a wide range of the parameter space \cite{MG02} .

The analysis to first order in $\epsilon$ is similar to the one worked out 
in Ref.\ \onlinecite{GD02} for the free cooling case. Some details of the 
derivation of the transport coefficients as well as their final 
expressions are given in Appendix \ref{appA}. Here, given that the 
theoretical predictions for the shear viscosity coefficient $\eta$ will be 
compared with those obtained from Monte Carlo simulations, we focus 
our attention on the explicit final expression for $\eta$. According to Eq.\ 
(\ref{a25}), the shear viscosity can be written as 
\begin{equation}
\label{3.5}
\eta=\frac{nT}{\nu}\eta^*,
\end{equation}
where 
\begin{equation}
\label{3.5.1}
\nu=\sqrt{\pi}n\sigma_{12}^2v_0, 
\end{equation}
is an effective collision frequency and the reduced shear viscosity coefficient $\eta^*$ is 
\begin{equation}
\label{3.5.2}
\eta^*=x_1\gamma_1^2d_{1,1}^*+x_2\gamma_2^2d_{2,1}^*
\end{equation}
with 
\begin{equation}
\label{3.6}
d_{1,1}^*=\frac{\gamma_2(\tau_{22}^*-\zeta^*)-\gamma_1\tau_{12}^*}{\Delta},
\quad d_{2,1}^*=\frac{\gamma_1(\tau_{11}^*-\zeta^*)-\gamma_2\tau_{21}^*}
{\Delta}.
\end{equation}
Here, $\zeta^*=\zeta^{(0)}/\nu$, 
\begin{equation}
\label{3.7}
\Delta=\gamma_1\gamma_2\left[\zeta^{*2}-\zeta^*(\tau_{11}^*+\tau_{22}^*)+
\tau_{11}^*\tau_{22}^*-\tau_{12}^*\tau_{21}^*\right], 
\end{equation}
and the dimensionless quantities $\tau_{ij}^*$ are given by \cite{GD02}
\begin{eqnarray}
\label{3.8}
\tau_{11}^*&=&\frac{16}{5\sqrt{2}}x_1\left(\frac{\sigma_{1}}
{\sigma_{12}}\right)^2
\theta_1^{-1/2}\left[1-\frac{1}{4}(1-\alpha_{11})^2\right]
\left(1-\frac{c_1}{64}
\right)\nonumber\\
& & +\frac{8}{15}x_2\mu_{21}(1+\alpha_{12})\theta_1^{3/2}\theta_2^{-1/2}
\left[6\theta_1^{-2}(\mu_{12}\theta_2-\mu_{21}\theta_1)
(\theta_1+\theta_2)^{-1/2}\right.\nonumber\\
& & +\frac{3}{2}\mu_{21}\theta_1^{-2}(\theta_1+\theta_2)^{1/2}(3-\alpha_{12})
+5\theta_1^{-1}(\theta_1+\theta_2)^{-1/2}\nonumber\\
& & \left. +\frac{c_2}{16}\frac{2\theta_2(12\mu_{21}+9\mu_{12}-10)-
\theta_1(5-6\mu_{21})-\case{3}{2}\mu_{21}(3-\alpha_{12})(\theta_1+\theta_2)}
{(\theta_1+\theta_2)^{5/2}}\right],
\end{eqnarray}
\begin{eqnarray}
\label{3.9}
\tau_{12}^*&=&\frac{8}{15}x_2\frac{\mu_{21}^2}{\mu_{12}}(1+\alpha_{12})
\theta_1^{3/2}\theta_2^{-1/2}
\left[6\theta_2^{-2}(\mu_{12}\theta_2-\mu_{21}\theta_1)
(\theta_1+\theta_2)^{-1/2}\right.\nonumber\\
& & +\frac{3}{2}\mu_{21}\theta_2^{-2}(\theta_1+\theta_2)^{1/2}(3-\alpha_{12})
-5\theta_2^{-1}(\theta_1+\theta_2)^{-1/2}\nonumber\\
& & \left. +\frac{c_1}{16}\frac{2\theta_1(10-12\mu_{12}-9\mu_{21})+
\theta_2(5-6\mu_{12})-\case{3}{2}\mu_{21}(3-\alpha_{12})(\theta_1+\theta_2)}
{(\theta_1+\theta_2)^{5/2}}\right].
\end{eqnarray}
The corresponding expressions for $\tau_{22}^*$ and $\tau_{21}^*$ can be 
inferred from Eqs.\ (\ref{3.8}) and (\ref{3.9}) by interchanging 
$1\leftrightarrow2$.

Equations (\ref{3.5})--(\ref{3.9}) provide the explicit expression for the 
shear viscosity $\eta$ of a heated granular binary mixture in the first 
Sonine approximation. It is apparent that the reduced viscosity $\eta^*$ 
presents a complex nonlinear dependence on the restitution coefficients 
$\alpha_{11}$, $\alpha_{22}$, and $\alpha_{12}$ and the parameters of the mixture 
$m_1/m_2$, $\sigma_1/\sigma_2$, and $n_1/n_2$.  The quality of the expression for $\eta^*$ will be 
later assessed by comparison with Monte Carlo simulations in the uniform 
shear flow (USF) problem with a thermostat. Before studying 
the general dependence of $\eta^*$ on the parameter space, it is 
instructive to consider some special limit cases. In the elastic limit, 
$\alpha_{11}=\alpha_{22}=\alpha_{12}=1$, $\zeta^*=0$, $\gamma_i=1$,
and $c_1=c_2=0$, so that the expression (\ref{3.5.2}) becomes
\begin{equation}
\label{3.10}
\eta^*=\frac{x_1^2R_1+x_2^2R_2+x_1x_2R_{12}}
{x_1^2S_1+x_2^2S_2+x_1x_2S_{12}},
\end{equation}
where
\begin{equation}
\label{3.11}
R_1=\frac{2}{3}+\frac{2\mu}{5}, \quad R_2=\frac{2}{3}+\frac{2}{5\mu},
\end{equation}
\begin{equation}
\label{3.12}
R_{12}=\frac{8}{15}+\frac{\sqrt{2}}{5}\left[\left(\frac{\sigma_{1}}{\sigma_{12}}\right)^2
\mu_{21}^{-1/2}\mu_{12}^{-1}+\left(\frac{\sigma_{2}}{\sigma_{12}}\right)^2
\mu_{12}^{-1/2}\mu_{21}^{-1}\right], 
\end{equation}
\begin{equation}
\label{3.13}
S_1=\frac{16}{5\sqrt{2}}R_1\left(\frac{\sigma_{1}}{\sigma_{12}}\right)^2
\mu_{21}^{1/2}, \quad 
S_2=\frac{16}{5\sqrt{2}}R_2\left(\frac{\sigma_{2}}{\sigma_{12}}\right)^2
\mu_{12}^{1/2},
\end{equation}
\begin{equation}
\label{3.14}
S_{12}=\frac{32}{15}+\frac{16}{25}\left(\frac{\sigma_{1}}{\sigma_{12}}\right)^2
\left(\frac{\sigma_{2}}{\sigma_{12}}\right)^2
\left(\mu_{12}\mu_{21}\right)^{-1/2}. 
\end{equation}
Here, $\mu=\mu_{12}/\mu_{21}=m_1/m_2$ is the mass ratio. Equation 
(\ref{3.10}) agrees with the results obtained in the first 
Sonine approximation to the coefficient of viscosity of a molecular 
gas-mixture of hard spheres \cite{note}. In the case of mechanically 
equivalent particles ($m_1=m_2$, 
$\alpha_{11}=\alpha_{22}=\alpha_{12}\equiv\alpha$, $\sigma_1=\sigma_2
\equiv\sigma$), $\gamma_i=1$, $\zeta^*=(2/3)(1-\alpha^2)(1+3c/32)$, and 
\begin{equation}
\label{3.15}
c_1=c_2=c=\frac{32(1-\alpha)(1-2\alpha^2)}{81
-17\alpha+30\alpha^2(1-\alpha)}
\end{equation}
In this case, one gets
\begin{equation}
\label{3.16}
\eta^*=\frac{15}{4}\left((1+\alpha)\left[(2+\alpha)+
\frac{c}{128}(33\alpha-39)\right]\right)^{-1}.
\end{equation}
This expression coincides with the one recently obtained for a heated 
granular monocomponent gas \cite{GM02}. All this shows the self-consistency 
of the present description.  

\section{Monte Carlo simulation for uniform shear flow with thermostat}
\label{sec4}

The USF is a nonequilibrium state characterized by constant partial 
densities $n_i$, a linear velocity profile ${\bf u}={\bf u}_i={\sf 
a}\cdot{\bf r}$, where the elements of the tensor ${\sf a}$ are $a_{k\ell}=a
\delta_{kx}\delta_{\ell y}$, $a$ being the constant shear rate. In addition, 
the granular temperature $T$ and the pressure tensor ${\sf P}$ are uniform, 
while the mass and heat fluxes vanish by symmetry reasons. 
This special state is generated by Lees-Edwards boundary 
conditions\cite{LE72} which are simple boundary periodic boundary conditions 
in the local Lagrangian frame ${\bf R}\equiv {\bf r}-{\sf a}\cdot {\bf r}t$ 
and ${\bf V}\equiv {\bf v}-{\sf a}\cdot {\bf r}$. In terms of these 
variables the velocity distribution functions are uniform\cite{DSBR86}
\begin{equation}
\label{4.0}
f_i({\bf r}, {\bf v};t)=f_i({\bf V};t).
\end{equation}

In the case of elastic collisions ($\zeta=0$) and in the absence of a 
thermostatting force, the energy balance equation (\ref{2.11}) yields the 
heating equation
\begin{equation}
\label{4.1}
\partial_tT=-\frac{2}{3n}aP_{xy}. 
\end{equation}
Since the granular temperature $T$ increases in time, so does 
the collision frequency $\nu(t)$ according to Eq.\ (\ref{3.5.1}). As a 
consequence, the reduced shear rate $a^*(t)=a/\nu(t)$ (which is the relevant 
uniformity parameter) monotonically decreases with time and the system 
asymptotically tends towards that of (local) equilibrium. 
This implies that for sufficiently long times (which means here $a^*\ll 1$) 
the system reaches a regime described by linear hydrodynamics and the 
Navier-Stokes shear viscosity $\eta$ can be identified as\cite{NO79,MS96} 
\begin{equation}
\label{4.2}
\frac{\nu}{nT}\eta=-\lim_{t\to \infty}\frac{P_{xy}^*}{a^*},
\end{equation}
where $P_{xy}^*=P_{xy}/nT$. This route has been shown to be quite 
efficient to measure the Navier-Stokes shear viscosity 
coefficient for dilute\cite{Pepe} and dense\cite{MS96} gases.

For a granular mixture, unless a thermostat is introduced, 
the energy balance equation (\ref{2.11}) leads to a steady state when the 
viscous heating effect is exactly balanced by the collisional cooling\cite{MG02a}. 
However, if the granular mixture is excited by the Gaussian thermostat 
\begin{equation}
\label{4.3}
{\bf F}_i^{\text{th}}=\frac{1}{2}m_i\zeta{\bf V}, 
\end{equation}
that exactly compensates for the collisional energy loss, the 
viscous heating still heats the system and Eq.\ (\ref{4.1}) 
remains valid. Consequently, the linear relationship (\ref{4.2}) allows one 
to determine the shear viscosity coefficient in the long time 
limit. Recently, this idea has been used to measure the shear viscosity of a 
heated granular monocomponent gas\cite{GM02}. The comparison with 
kinetic theory shows an excellent agreement over a wide range of values of 
the restitution coefficient. It must be noted that here $\eta$ represents 
the shear viscosity of an {\em excited} granular mixture and thus it does 
not necessarily coincide with the Navier-Stokes shear viscosity obtained in 
the unforced case\cite{GD02}. As a matter of fact, the results obtained in 
Sec.\ \ref{sec3} indicate that the transport properties are affected by the 
Gaussian thermostat and the expression (\ref{3.5.2}) for the (reduced) shear 
viscosity differs from the one derived in the freely cooling 
case\cite{GD02}.  The use of thermostats to control collisional cooling in 
undriven systems is quite common\cite{BT02,variosbis}. Usually, 
the motivation is to produce a steady state while here is to remove the 
steady state in favor of one whose dynamics determines the viscosity.

The Boltzmann equation for a mixture of inelastic hard spheres under 
USF and subject to the external Gaussian force 
(\ref{4.3}) reads: 
\begin{equation} 
\partial _{t}f_i-aV_y\frac{\partial}{\partial V_x} 
f_{i}+\frac{1}{2}\zeta
\frac{\partial}{\partial {\bf V}}\cdot \left({\bf V}f_i\right)
=\sum_{j}J_{ij}\left[ {\bf V}|f_{i}(t),f_{j}(t)\right] \;.
\label{4.4} 
\end{equation} 
The second term on the left-hand side represents an inertial force of the 
form ${\bf F}_i^{\text{in}}=-m_i{\sf a}\cdot {\bf V}$, while the third term 
represents the thermostat force ${\bf F}_i^{\text{th}}$ given 
by Eq.\ (\ref{4.3}). Thus, in this frame, the system is in a homogenous state 
subjected to the action of the (total) force ${\bf F}_i^{\text{in}}+
{\bf F}_i^{\text{th}}$. We have numerically solved Eq.\ (\ref{4.4}) by means 
of the Direct Simulation Monte Carlo (DSMC) method\cite{B94}. This method 
was devised to mimic the processes involved in the Boltzmann collision term 
and its extension to deal with inelastic collisions is straightforward. In 
addition, since the USF is spatially homogeneous in the Lagrangian 
frame, the simulation method is easy 
to carry out and only the (peculiar) velocities of the particles need to be stored. 
The restriction to this homogeneous state prevents us from studying the 
possible formation of particle clusters (microstructure).

Technical details of the DSMC method and its application to the USF state can 
be found in Refs.\ \cite{MG02} and \cite{MG02a}.  In our simulations we have typically taken 
a total number of particles  $N=10^5$, a number of replicas 
${\cal N}=5$, and a time step $\Delta t=3 \times 10^{-3} \ell_{11}/v_{01}$, where 
$\ell_{11}=\left(\sqrt{2}\pi n_1\sigma_1^2\right)^{-1}$ is the mean free path for 
collisions $1-1$ and $v_{01}=\sqrt{2T/m_1}$.

At given values of the shear rate $a$, the restitution coefficients 
$\alpha_{ij}$, and the parameters of the mixture, the system is 
initially prepared in a local equilibrium state with a temperature 
$T(0)=T_0$ such that the initial value of the reduced shear rate is 
$a_0^*=a/\nu(T_0)$. As the system evolves, we monitor the time 
evolution of the reduced shear rate $a^*(t)=a/\nu(T(t))$
and the reduced  $xy$ element of the pressure tensor
$P_{xy}^*(t)=P_{xy}(t)/nT(t)$. We observe
that in all the cases, after a transient period, the ratio 
$\eta^*\equiv -P_{xy}^*/a^*$ reaches a constant value that is independent of 
the shear rate and time. This allows us to measure the corresponding 
shear viscosity coefficient $\eta$ as  
\begin{equation}
\label{4.5}
\eta(t)=\frac{nT(t)}{\nu(t)}\eta^*,
\end{equation}
where the dimensionless shear viscosity $\eta^*$ is independent of 
time but depends on dissipation and the parameters of the mixture 
(masses, sizes and concentrations).

The theoretical prediction for $\eta^*$ can be obtained from the 
Chapman-Enskog solution to Eq.\ (\ref{4.4}) up to first order in the shear 
rate $a$. In Appendix \ref{appB} it 
is easily proved that the first order solution to (\ref{4.4}) leads 
to the same expression for 
the shear viscosity as the one obtained in Sec.\ \ref{sec3} from the general 
Chapman-Enskog method specialized to USF. Thus, in the first Sonine 
approximation, the theoretical prediction of $\eta^*$ is given by Eqs.\ 
(\ref{3.5.2})--(\ref{3.9}).

Before analyzing the dependence of the dimensionless shear 
viscosity coefficient $\eta^*$ on the parameters of the problem, it is 
instructive to test the consistency of the simulation method in the limit 
$a^*\to 0$ (which corresponds here to $t\nu \gg 1$) . 
For long times and for given values of $\alpha$, $m_1/m_2$, 
$\sigma_1/\sigma_2$ , and $n_1/n_2$, the reduced viscosity $\eta^*$ 
must reach a value independent of the inital preparation of the 
system. In Fig.\ \ref{fig1}, we plot the shear-rate dependent viscosities 
$\eta^*(a^*)$ measured in the simulation, relative to its Navier-Stokes 
value $\eta^*_{\text{B}}$ given by the Boltzmann theory [Eqs.\ 
(\ref{3.5.2})--(\ref{3.9})] for three different choices of the initial shear rate 
$a_0^*$: 0.2, 0.3, and 0.4.  Here, the restitution coefficient is $\alpha=0.9$, the mass ratio 
is $m_1/m_2=4$, the concentration ratio is $n_1/n_2=1$ and the size ratio 
is $\sigma_1/\sigma_2=3$. After a transient regime of a few mean free times, 
we observe that the curves corresponding to the three different initial 
conditions practically coincide. This means that a hydrodynamic regime
independent of the inital conditions has been achieved. In 
addition, for very small values of $a^{*2}$, the ratio $\eta^*(a^*)/
\eta^*_{\text{B}}$ fluctuate around 1 showing that in this regime the 
viscosity coefficient measured in the simulation is consistent with the 
value obtained from the Boltzmann kinetic theory. The same behavior has been 
found for other values of the restitution coefficient as well as of the 
parameters of the mixture. Notice that the limit $a^*\to 0$ is strictly 
unattainable in the USF because it requires an infinite amount of time. 
Also, the signal-to-noise ratio decreases in that limit so that the 
fluctuations increase.

\section{Comparison between theory and simulation}
\label{sec5}

Once the consistency of the simulation method has been tested, we focus  
our attention on the study of transport properties in the Navier-Stokes regime. 
In this Section we compare the predictions of 
the Sonine  approximation with the results obtained from the DSMC method.  
A complete presentation of the results is complex due to the high dimensionality of 
the parameter space: $\{\alpha_{11}, \alpha_{22}, \alpha_{12}, 
m_1/m_2, \sigma_1/\sigma_2, n_1/n_2\}$. For 
the sake of concreteness, henceforth we will assume that the spheres are 
made of the same material, i.e., $\alpha_{11}=\alpha_{22}=\alpha_{12}\equiv 
\alpha$. This reduces the parameter space to four quantities.

Apart from the shear viscosity coefficient, another interesting quantity at 
this level of description is the temperature ratio $T_1/T_2$. This ratio 
measures the breakdown of the energy equipartition. The analysis of the 
temperature differences has been a subject of growing interest in the 
past few years among both theorists \cite{DHGD02,MG02,BT02,GD99b} and 
experimentalists\cite{WP02,FM02}. As was previously found from the Boltzmann 
kinetic theory\cite{GD99b}, except for mechanically equivalent particles, 
the partial temperatures $T_i$ are different. For the sake of 
illustration, Fig.\  \ref{fig2} shows the dependence of the temperature 
ratio on the size ratio $\sigma_1/\sigma_2$ for an equimolar mixture 
($n_1/n_2=1$) and three different values of the restitution coefficient 
$\alpha=0.9, 0.8$, and $0.7$. We consider a binary mixture of constant 
density and so, $m_1/m_2=(\sigma_1/\sigma_2)^3$. We observe that for large 
size ratios the temperature differences are quite important, 
even for moderate dissipation. It is also apparent that an excellent 
agreement between the theory (given by the first Sonine correction) and 
Monte Carlo simulations (symbols) is found over the entire range of values 
of size and mass ratios considered.

Next, we explore the influence of dissipation on the reduced shear viscosity 
$\eta^*(\alpha)$ for different values of the mass ratio, the size ratio, and 
the concentration ratio. Three different values of the (common) restitution coefficient are 
considered: $\alpha=0.9, 0.8$, and $0.7$. In Fig.\ \ref{fig3}, we plot the 
ratio $\eta^*(\alpha)/\eta^*(1)$ versus the mass ratio $m_1/m_2$ for 
$\sigma_1/\sigma_2=n_1/n_2=1$. Here, $\eta^*(1)$ refers to the elastic value 
for the shear viscosity coefficient. Again, the symbols represent the 
simulation data while the lines refer to the theoretical results obtained 
from the Boltzmann equation in the first Sonine approximation. We see that 
in general the deviation of $\eta^*(\alpha)$ from its functional form for 
elastic collisions is quite important. This tendency becomes more 
significant as the mass disparity increases. The agreement between the first 
Sonine approximation and simulation is seen to be in general excellent. 
This agreement is similar to the one previously found in the monocomponent 
case\cite{GM02}. At a quantitative level, the discrepancies between theory 
and simulation tend to increase as the restitution coefficient decreases, 
although these differences are quite small (say, for instance, around 2\% 
at $\alpha=0.7$ in the disparate mass case $m_1/m_2=10$).

The influence of the size ratio on the shear viscosity is shown in Fig.\ 
\ref{fig4} for $m_1/m_2=4$ and $n_1/n_2=1$. We observe again a strong 
dependence of the shear viscosity on dissipation. However, for a given value 
of $\alpha$, the influence of $\sigma_1/\sigma_2$ on $\eta^*$ is weaker than 
the one found before in Fig.\ \ref{fig3} for the mass ratio. 
The agreement for both $\alpha=0.9$ and $\alpha=0.8$ is quite good, except 
for the largest size ratio at $\alpha=0.8$. These discrepancies become more 
significant as the dissipation increases (say  $\alpha=0.7$),  
especially for mixtures of particles of very different sizes. Finally, 
Fig.\ \ref{fig5} shows the dependence of $\eta^*(\alpha)/\eta^*(1)$ on 
the concentration ratio for $m_1/m_2=4$ and $\sigma_1/\sigma_2=1$. We 
observe that both the theory and simulation predicts a very weak influence 
of composition on the shear viscosity. With respect to the influence of 
dissipation, the trends are similar to those of Figs.\ \ref{fig3} and 
\ref{fig4}: the main effect of inelasticity in collisions is to enhance the 
momentum transport with respect to the case of elastic collisions. The 
agreement now between theory and simulation is very good, even for disparate 
values of the concentration ratio and/or strong dissipation.  
Therefore, according to the comparison carried out in Figs.\ 
\ref{fig3}, \ref{fig4}, and \ref{fig5}, we can conclude that the agreement 
extends over a wide range values of the restitution coefficient, indicating 
the reliability of the first Sonine approximation for describing granular 
flows beyond the quasielastic limit.

\section{Discussion}
\label{sec6}

Although the utility of a hydrodynamic description for granular media under 
rapid flow conditions has been recognized for many years, its
domain of validity as well as
the forms of the transport coefficients remain a topic of interest and 
controversy. In this context, there are some doubts about the possibility of 
going from a kinetic theory to a hydrodynamic level of description by using 
a Chapman-Enskog expansion around the homogenous cooling state.
Given that the search for exact solutions of the Boltzmann equation
is far beyond the present perspectives,
an alternative to get some insight into
the above question is to numerically solve the kinetic equation and compare
these results with the corresponding solution obtained by assuming the validity of
a hydrodynamic description. In this paper, we have performed
such a comparison at the level of the shear viscosity coefficient of a
heated granular mixture. The system is heated by the action of a
thermostatting external force which exactly compensates for cooling effects
associated with the inelasticity of collisions.
Although some previous works \cite{GM02,BRC99,BSSS99} have compared kinetic
theory predictions for transport coefficients with computer simulations in the
case of a monocomponent gas, studies for multicomponent granular gases are more scarce. Very recently, a seemingly similar analysis for the shear viscosity $\eta$ of a dense mixture has been given in Ref.\ \onlinecite{AWAL02}. Nevertheless, the above kinetic theory only holds for nearly elastic particles and the expression of $\eta$ in the first Sonine approximation coincides with the one obtained in the elastic case.

As a first step in our issue, in Sec.\ \ref{sec3}
we have derived the general hydrodynamic equations
of a {\em heated} binary mixture of smooth inelastic spheres from the
Boltzmann kinetic equation by using the Chapman-Enskog method. The
corresponding transport coefficients have been expressed
in terms of the solution to integral equations, which are then solved 
approximately (first Sonine polynomial approximation) just as in the case of 
elastic collisions.  The explicit expressions for 
the set of relevant transport coefficients $\{D, D_p, D', D'', L, 
\lambda, \eta\}$ are displayed in Appendix \ref{appA}.  In contrast to previous works\cite{varios,AWAL02}, our
results are not limited a priori to weak inelasticity and they take into account 
the effect of the temperature differences on the transport coefficients. 
On the other hand, the results obtained here for the
transport coefficients slightly differ from those obtained in the freely 
cooling case \cite{GD02}, showing that in general the introduction
of a thermostat affects the transport properties of the gas \cite{DSBR86}.
The Chapman-Enskog results obtained for the mixture have been then
specialized to the hydrodynamic state of
transverse shear. In this state, the shear viscosity coefficient
$\eta$ is the relevant transport coefficient of the problem. The explicit
form of $\eta$ is given by Eqs.\ (\ref{3.5.2})--(\ref{3.9})
in terms of the restitution coefficients $\alpha_{ij}$
and the parameters of the mixture (masses, diameters, and concentrations).

To test the assumptions of the Chapman-Enskog method and the approximate
Sonine solution to the resulting integral equation, the DSMC method has been 
used to solve the Boltzmann equation in the uniform shear flow state.
In the absence of a thermostat, in a granular fluid there is a
competition between two opposite effects:
viscous heating and collisional cooling. In that case, when both effects
exactly cancel each other, a steady state is reached after a transient
period.  In this steady state, due to the coupling between dissipation and 
the shear rate,  the system is far away from the Navier-Stokes regime, 
except when $\alpha\to 1$ \cite{MG02a}. 
However, if the external thermostat is adjusted to compensate for
the energy lost in collisions, the shearing work still heats the system.
As a consequence, as the system evolves, the reduced shear rate $a^*(t)$
goes to zero and the system achieves a regime described by linear
hydrodynamics. In this regime, the Navier-Stokes shear viscosity coefficient
can be measured from simulations. In this paper, the 
thermostat is used to remove the steady state
in favor of a time-dependent state
whose dynamics allows one to get the Navier-Stokes shear viscosity just as 
for the case of elastic collisions\cite{MS96,Pepe}.

The dependence of the viscosity $\eta$ on the full parameter 
space has been explored. Specifically, the parameter space over which
our solution has been verified is the mass ratio $m_1/m_2$, the concentration
ratio $n_1/n_2$, the ratio of diameters $\sigma_1/\sigma_2$, and the (common)
restitution coefficient $\alpha\equiv \alpha_{11}=\alpha_{11}=\alpha_{22}
=\alpha_{12}$. The theory and simulation clearly show how in general,
the influence of dissipation on momentum transport is quite important
since there is a
relevant dependence of the viscosity $\eta(\alpha)$
on the restitution coefficient $\alpha$.
At a given value of the restitution coefficient, the dependence of 
$\eta(\alpha)/\eta(1)$ on the mass ratio is more significant
than the one found on the composition and diameters.
This feature has been also found for the temperature ratio in 
the experiments recently carried out in vibrated mixtures \cite{WP02,FM02},
although experimental confirmation of the trends observed here
for the viscosity is still lacking. With respect to the accuracy of the
theory predictions we see that, in general, the Chapman-Enskog results in the
first Sonine approximation exhibit an excellent agreement with the
simulation data. This supports the idea that
the Sonine polynomial approximation for granular fluids
has an accuracy comparable to that for elastic collisions.
Exceptions to this agreement are extreme mass ratios and
strong dissipation. These discrepancies are
due basically to the approximations introduced in
applying the Chapman-Enskog method, and more specifically in using
the first Sonine approximation.

One of the main limitations of the results obtained here from the 
Boltzmann equation is its restriction to the low-density regime. In 
this regime, the collisional transfer contributions to the fluxes are 
negligible and only their kinetic contributions are taken into 
account. Possible extension in both aspects, theory and 
simulation, of the present simple hydrodynamic state to higher 
densities can be carried out in the context of the revised Enskog 
theory. In this case, many of the phenomena appearing in dense 
granular fluids (such as spontaneous formation of 
dense clusters surrounded by regions of low-density \cite{GZ93}) 
could be studied. On the other hand, although the comparison 
performed here has been made undergoing {\em uniform} shear flow 
without paying attention to the possible formation of particle 
clusters\cite{clusters}, our Chapman-Enskog results
apply for general {\em inhomogeneous} situations. The only 
restriction is that they provide the irreversible parts of the mass, 
heat, and momentum fluxes to leading order in the spatial 
gradients of the hydrodynamic fields. In this context,  the results derived in 
this paper can be used to analyze the behavior of granular mixtures 
in a lot of physical situations. Thus, for instance, the knowledge of 
the complete hydrodynamic equations for a  binary mixture allows 
one to say whether the mixture hydrodynamics is more or less 
unstable to long-wavelength perturbations than that of the 
one-component case, and  what are the mechanisms involved in 
phenomena very often observed in nature and experiments such as 
phase separation or {\em segregation}.  We hope that the present 
results give some insight into the understanding of these 
interesting and complex problems.

\acknowledgments 

V. G. acknowledges partial support from the Ministerio de Ciencia y
Tecnolog\'{\i}a (Spain) through Grant No. BFM2001-0718.

\appendix
\section{Chapman-Enskog expansion}
\label{appA}

In this Appendix, the expressions of the transport coefficients for a heated 
granular mixture are obtained. The derivation follows similar steps as those 
made in Ref.\ \cite{GD02} in the free cooling case. Here, we will use 
the same notation as in Ref.\ \cite{GD02}. In the first order, the 
distribution function $f_i^{(1)}$ verifies the kinetic equation
\begin{equation}
\label{a1} 
\left({\cal L}_i+\frac{1}{2}\zeta^{(0)}\frac{\partial}{\partial {\bf V}}
\cdot {\bf V}\right)f_i^{(1)}+{\cal M}_if_j^{(1)}=
-\left(D_t^{(1)}+{\bf V}\cdot \nabla\right)f_i^{(1)},
\end{equation}
where $D_t^{(1)}=\partial_t^{(1)}+{\bf u}\cdot \nabla$, and 
\begin{equation}
\label{a2}
{\cal L}_if_i^{(1)}=-\left(J_{ii}[f_i^{(0)},f_i^{(1)}]+
J_{ii}[f_i^{(1)},f_i^{(0)}]+J_{ij}[f_i^{(1)},f_j^{(0)}]\right), 
\end{equation}
\begin{equation}
\label{a3}
{\cal M}_if_j^{(1)}=-J_{ij}[f_i^{(0)},f_j^{(1)}]. 
\end{equation}
In these equations, it is understood that $i\neq j$ and use has been made of 
the fact that $\partial_t^{(0)}T=0$ and the results ${\bf j}_i^{(0)}=
{\bf q}^{(0)}=\zeta^{(1)}=0$. The last equality follows from the fact that 
the cooling rate is a scalar, and so $\zeta^{(1)}$ should be proportional 
to $\nabla \cdot {\bf u}$. However, as  shown later, there is no 
contribution to $f_i^{(1)}$ proportional to the divergence of the flow 
field so that $\zeta^{(1)}=0$ by symmetry. This property is special to the 
low density Boltzmann kinetic theory and such terms occur at higher 
densities \cite{GD99a}. The macroscopic balance equations to first order 
are 
\begin{equation}
\label{a4}
D_t^{(1)}x_1=0,\quad \frac{3}{5}D_t^{(1)}\ln p=\frac{3}{2}D_t^{(1)}\ln T=
-\nabla \cdot {\bf u},\quad D_t^{(1)}{\bf u}=-\rho^{-1}\nabla p.
\end{equation}
Use of these in (\ref{a1}) yields 
\begin{equation}
\label{a5}
\left({\cal L}_i+\frac{1}{2}\zeta^{(0)}\frac{\partial}{\partial {\bf V}}
\cdot {\bf V}\right)f_i^{(1)}+{\cal M}_if_j^{(1)}=   
{\bf A}_i\cdot \nabla x_1+
{\bf B}_i\cdot \nabla p+{\bf C}_i\cdot \nabla 
T+D_{i,\alpha\beta}\nabla_{\alpha}u_{\beta},
\end{equation} 
where 
\begin{equation}
{\bf A}_{i}({\bf V})=-\left( \frac{\partial }{\partial x_{1}}
f_{i}^{(0)}\right) _{p,T}{\bf V},  \label{a6}
\end{equation}
\begin{equation}
{\bf B}_{i}({\bf V})=-\frac{1}{p}\left[ f_{i}^{(0)}{\bf V}+\frac{nT}{\rho }
\left( \frac{\partial }{\partial {\bf V}}f_{i}^{(0)}\right) \right],
\label{a7}
\end{equation}
\begin{equation}
{\bf C}_{i}({\bf V})=\frac{1}{T}\left[ f_{i}^{(0)}+\frac{1}{2}\frac{
\partial }{\partial {\bf V}}\cdot \left( {\bf V}f_{i}^{(0)}\right) \right] 
{\bf V},
\label{a8}
\end{equation}
\begin{equation}
D_{i,\alpha \beta }({\bf V})=V_{\alpha }\frac{\partial }{\partial V_{\beta }}
f_{i}^{(0)}-\frac{1}{3}\delta _{\alpha \beta }{\bf V}\cdot \frac{\partial }{
\partial {\bf V}}f_{i}^{(0)}.  
\label{a9}
\end{equation}
The solutions to Eqs.\ (\ref{a5}) are of the form 
\begin{equation}
f_{i}^{(1)}={\sf {\cal A}}_{i}\cdot \nabla x_{1}+{\sf {\cal B}}_{i}\cdot
\nabla p+{\sf {\cal C}}_{i}\cdot \nabla T+{\sf {\cal D}}_{i,\alpha \beta
}\nabla_{\alpha }u_{\beta }\;.  \label{a10}
\end{equation}
The coefficients ${\sf {\cal A}}_{i},{\sf {\cal B}}_{i},{\sf {\cal C}}_{i},$
and ${\sf {\cal D}}_{i,\alpha \beta }$ are functions of the peculiar
velocity ${\bf V}$ and the hydrodynamic fields. The cooling rate depends
on space through its dependence on $x_{1}$, $p$, and $T$. 
The integral equations for the unknowns are easily identified as 
coefficients of the independent gradients in (\ref{a10}). The result is 
\begin{equation}
\label{a11}
\left({\cal L}_i+\frac{1}{2}\zeta^{(0)}\frac{\partial}{\partial {\bf V}}
\cdot {\bf V}\right)
\left(
\begin{array}{c}
{\sf {\cal A}}_i\\
{\sf {\cal B}}_i\\
{\sf {\cal C}}_i\\
{\sf {\cal D}}_{i,\alpha\beta}
\end{array}
\right)+
{\cal M}_i
\left(
\begin{array}{c}
{\sf {\cal A}}_j\\
{\sf {\cal B}}_j\\
{\sf {\cal C}}_j\\
{\sf {\cal D}}_{j,\alpha\beta}
\end{array}
\right)
=
\left(
\begin{array}{c}
{\bf A}_i\\
{\bf B}_i\\
{\bf C}_i\\
{\bf D}_{i,\alpha\beta}
\end{array}
\right).
\end{equation}
Note that, in contrast to what happens in the free cooling case\cite{GD02}, 
here each one of the quantities ${\sf {\cal A}}_i$, ${\sf {\cal B}}_i$, 
${\sf {\cal C}}_i$, and ${\sf {\cal D}}_{i,\alpha\beta}$ obey closed 
integral equations. The solution to Eq.\ (\ref{a11}) provides the expression for 
the transport coefficients. In the case of the mass flux ${\bf j}_1$, these 
coefficients are identified as 
\begin{equation}
D=-\frac{\rho }{3m_{2}n}\int d{\bf v}\,{\bf V}{\bf \cdot \,}{\cal A}
_{1},  \label{a12}
\end{equation}
\begin{equation}
D_{p}=-\frac{m_{1}p}{3\rho}\int d{\bf v}\,{\bf V}\,\cdot {\sf {\cal 
B}}_{1} , \label{a13}
\end{equation}
\begin{equation}
D^{\prime }=-\frac{m_{1}T}{3\rho }\int d{\bf v}\,{\bf V}\,\cdot {\sf 
{\cal C}}_{1} . \label{a14}
\end{equation}
The transport coefficients for the heat flux are 
\begin{equation}
D^{''}=-\frac{1}{3T^{2}}\sum_{i=1}^2\,\int d{\bf v}\,\frac{1}{2}
m_{i}V^{2}{\bf V}\,\cdot {\sf {\cal A}}_{i},  \label{a15}
\end{equation}
\begin{equation}
L=-\frac{1}{3}\sum_{i=1}^2\,\int d{\bf v}\,\frac{1}{2}
m_{i}V^{2}{\bf V}\cdot \,{\sf {\cal B}}_{i},  \label{a16}
\end{equation}
\begin{equation}
\lambda =-\frac{1}{3}\sum_{i=1}^2\,\int d{\bf v}\,\frac{1}{2}m_{i}V^{2}
{\bf V}\,\cdot {\sf {\cal C}}_{i}.  \label{a17}
\end{equation}
Finally, the shear viscosity is given by  
\begin{equation}
\eta =-\frac{1}{10}\sum_{i=1}^2\,\int d{\bf v}\,
m_{i}V_{\alpha}V_{\beta}{\sf {\cal D}}_{i,\alpha\beta}.  \label{a18}
\end{equation}

Accurate aproximations to the solutions to the integral equations for 
$\left({\sf {\cal A}}_i, {\sf {\cal B}}_i, 
{\sf {\cal C}}_i, {\sf {\cal D}}_{i,\alpha\beta}\right)$ may be obtained 
using low order truncation of expansions in a series of Sonine polynomials. 
In the case of the mass flux, we consider the leading Sonine approximation 
(lowest degree polynomial)
\begin{equation}
\label{a19}
\{{\sf {\cal A}}_i, {\sf {\cal B}}_i, {\sf {\cal C}}_i\}\to f_{i,M}{\bf 
V}\{a_{i,1}, b_{i,1}, c_{i,1}\},\quad 
f_{i,M}({\bf V})=n_i(m_i/2\pi T_i)^{3/2}\exp(-m_iV^2/2T_i),
\end{equation}
where $a_{i,1}=-(m_1m_2n/\rho n_iT_i)D$, $b_{i,1}=-(\rho/p n_iT_i)D_p$, and 
$c_{i,1}=-(\rho/T n_iT_i)D'$. The coefficients $a_{i,1}$, $b_{i,1}$, and 
$c_{i,1}$ are determined by multiplying the three first equations of Eq.\ 
(\ref{a11}) by $m_i{\bf V}$ and integrating over the velocity. The result is 
\begin{equation}
\label{a20}
a_{1,1}=-\left(\nu_{D}-\frac{1}{2}\zeta^{(0)}\right)^{-1}\left(\frac{\partial}
{\partial x_1}\ln n_1T_1\right)_{p,T},
\end{equation}
\begin{equation}
\label{a21}
b_{1,1}=-\left(\nu_{D}-\frac{1}{2}\zeta^{(0)}\right)^{-1}
\frac{n_1T_1}{p}\left(1-\frac{m_1nT}{\rho T_1}\right),
\end{equation}
\begin{equation}
\label{a22}
c_{1,1}=0.
\end{equation}
Here, the collision frequency $\nu_D$ is given by Eq.\ (73) of Ref.\ 
\cite{GD02}.

In the case of the pressure tensor, the leading Sonine approximation for the 
function ${\sf {\cal D}}_{i,\alpha\beta}$ is 
\begin{equation}
\label{a23}
{\sf {\cal D}}_{i,\alpha\beta}\to f_{i,M}d_{i,1}R_{i,\alpha\beta}, \quad  
R_{i,\alpha\beta}=m_i(V_{\alpha}V_{\beta}-\case{1}{3}V^2
\delta_{\alpha\beta}).
\end{equation}
The shear viscosity coefficient is given by 
\begin{equation}
\label{a24}
\eta=-nT^2\left(x_1\gamma_1^2d_{1,1}+x_2\gamma_2^2d_{2,1}\right).
\end{equation}
The coefficients $d_{i,1}$ can be determined by multiplying the fourth 
equation of Eq.\ (\ref{a11}) by $R_{i,\alpha\beta}$ and integrating over the 
velocity to get the coupled set of equations
\begin{equation}
\label{a25}
\left(
\begin{array}{cc}
\tau_{11}-\zeta^{(0)}&\tau_{12}\\
\tau_{21}&\tau_{22}-\zeta^{(0)}
\end{array}
\right)
\left(
\begin{array}{c}
d_{1,1}\\
d_{2,1}
\end{array}
\right)=-\left(
\begin{array}{c}
T_1^{-1}\\
T_2^{-1}
\end{array}
\right).
\end{equation}
The collision frequencies $\tau_{ij}=\tau_{ij}^*\nu$, where $\tau_{ij}^*$ 
are given by Eqs.\ (\ref{3.8}) and (\ref{3.9}). From Eq.\ (\ref{a25}) one 
easily gets the expression (\ref{3.5}) for the shear viscosity  given in the 
main text.

The calculations for the heat flux are similar to those previously made for 
the other fluxes. As in the unforced case, this requires going to the second 
Sonine approximation. In this case, the transport coefficients defining the 
heat flux (\ref{2.16}) are given by 
\begin{equation}
\label{a26}
D''=-\frac{5}{2}T\left(\frac{n_1\gamma_1^3}{m_1}a_{1,2}+
\frac{n_2\gamma_2^3}{m_2}a_{2,2}\right)+\frac{5}{2}\frac{nm_1m_2}{\rho T}
\left(\frac{\gamma_1}{m_1}-\frac{\gamma_2}{m_2}\right)D,
\end{equation}
\begin{equation}
\label{a27}
L=-\frac{5}{2}T^3\left(\frac{n_1\gamma_1^3}{m_1}b_{1,2}+
\frac{n_2\gamma_2^3}{m_2}b_{2,2}\right)+\frac{5}{2}\frac{\rho}{n}
\left(\frac{\gamma_1}{m_1}-\frac{\gamma_2}{m_2}\right)D_p,
\end{equation}
\begin{equation}
\label{a28}
\lambda=-\frac{5}{2}T^3\left(\frac{n_1\gamma_1^3}{m_1}c_{1,2}+
\frac{n_2\gamma_2^3}{m_2}c_{2,2}\right)+\frac{5}{2}\rho
\left(\frac{\gamma_1}{m_1}-\frac{\gamma_2}{m_2}\right)D'.
\end{equation}
The coefficients $a_{i,2}$, $b_{i,2}$ and $c_{i,2}$ obey the equations
\begin{equation}
\label{a29}
\left(
\begin{array}{cc}
\nu_{11}-\frac{3}{2}\zeta^{(0)}&\nu_{12}\\
\nu_{21}&\nu_{22}-\frac{3}{2}\zeta^{(0)}
\end{array}
\right)
\left(
\begin{array}{c}
a_{1,2}\\
a_{2,2}
\end{array}
\right)=\left(
\begin{array}{c}
X_1\\
X_2
\end{array}
\right), 
\end{equation}
\begin{equation}
\label{a30}
\left(
\begin{array}{cc}
\nu_{11}-\frac{3}{2}\zeta^{(0)}&\nu_{12}\\
\nu_{21}&\nu_{22}-\frac{3}{2}\zeta^{(0)}
\end{array}
\right)
\left(
\begin{array}{c}
b_{1,2}\\
b_{2,2}
\end{array}
\right)=\left(
\begin{array}{c}
Y_1\\
Y_2
\end{array}
\right), 
\end{equation}
\begin{equation}
\label{a31}
\left(
\begin{array}{cc}
\nu_{11}-\frac{3}{2}\zeta^{(0)}&\nu_{12}\\
\nu_{21}&\nu_{22}-\frac{3}{2}\zeta^{(0)}
\end{array}
\right)
\left(
\begin{array}{c}
c_{1,2}\\
c_{2,2}
\end{array}
\right)=\left(
\begin{array}{c}
Z_1\\
Z_2
\end{array}
\right), 
\end{equation}
where 
\begin{eqnarray}
\label{a32}
X_1&=& -\frac{\zeta^{(0)}m_1m_2nD}{\rho 
n_1T_1^2}-\frac{1}{2}\frac{T^2}{n_1T_1^3}\frac{\partial}{\partial x_1}\left(
n_1\gamma_1^2c_1\right)+\frac{2}{15}\frac{m_1^2m_2nD}{\rho n_1^2T_1^4}
\left[\int d{\bf v}_1 {\bf S}_1\cdot {\cal L}_1(f_{1,M}{\bf V}_1) \right.
\nonumber\\
& & \left. -\delta \gamma 
\int d{\bf v}_1 {\bf S}_1\cdot {\cal M}_1(f_{2,M}{\bf V}_2)\right],
\end{eqnarray}
\begin{eqnarray}
\label{a33}
Y_1&=&-\frac{\zeta^{(0)}\rho D_p}{p 
n_1T_1^2}-\frac{1}{2}\frac{c_1}{pT_1}
+\frac{2}{15}\frac{m_1\rho D_p}{p n_1^2T_1^4}
\left[\int d{\bf v}_1 {\bf S}_1\cdot {\cal L}_1(f_{1,M}{\bf V}_1) \right.
\nonumber\\
& & \left. -\delta \gamma 
\int d{\bf v}_1 {\bf S}_1\cdot {\cal M}_1(f_{2,M}{\bf V}_2)\right],
\end{eqnarray}
\begin{eqnarray}
\label{a34}
Z_1&=& -\frac{\zeta^{(0)}\rho D'}{T 
n_1T_1^2}-\frac{2+c_1}{2TT_1}
+\frac{2}{15}\frac{m_1\rho D'}{T n_1^2T_1^4}
\left[\int d{\bf v}_1 {\bf S}_1\cdot {\cal L}_1(f_{1,M}{\bf V}_1) \right.
\nonumber\\
& & \left. -\delta \gamma 
\int d{\bf v}_1 {\bf S}_1\cdot {\cal M}_1(f_{2,M}{\bf V}_2)\right],
\end{eqnarray}
and
\begin{equation}
\label{a35}
c_i=\frac{8}{15}\left[\frac{m_i^2}{4n_iT_i^2}\int d{\bf v}_1 V_1^4 
f_i^{(0)}-\frac{15}{4}\right].
\end{equation}
The corresponding expressions of the elements $X_2$, $Y_2$ and $Z_2$ can be 
deduced from Eqs.\ (\ref{a32}), (\ref{a33}) and (\ref{a34}), 
respectively, by interchanging $1\leftrightarrow2$ and setting $D\to D$, 
$D_p\to -D_p$ and $D'\to -D'$. The frequencies $\nu_{ij}$ and the collision 
integrals appearing in
Eqs.\ (\ref{a31}), (\ref{a32}), (\ref{a33}) and (\ref{a34})
were explicitly evaluated in
the Appendix D of Ref.\ \cite{GD02}. Thus, the transport coefficients 
$D''$, $L$, and $\lambda$ are completely determined.

\section{First order solution to the USF}
\label{appB}

In this Appendix we get the solution to Eq.\ (\ref{4.4}) in the first order 
in the shear rate $a$. The normal solution to Eq.\ (\ref{4.4}) is provided 
by the Chapman-Enskog method, i.e., a solution given as a power series in 
$a$:
\begin{equation}
\label{b1}
f_i=f_i^{(0)}+f_i^{(1)}+\cdots
\end{equation}
The zeroth-order solution $f_i^{(0)}$ verifies Eq.\ (\ref{3.3}) and it corresponds to 
the homogeneous cooling state distribution in the local Lagrangian frame. Its first 
Sonine approximation is given by Eq.\ (\ref{3.4}). Inserting the expansion (\ref{b1}) 
into Eq.\ (\ref{3.3}) leads to the following integral equation for $f_i^{(1)}$:
\begin{equation} 
\label{b2}
\partial _{t}f_i^{(1)}-aV_y\frac{\partial}{\partial V_x} 
f_{i}^{(0)}+\frac{1}{2}\zeta^{(0)}
\frac{\partial}{\partial {\bf V}}\cdot \left({\bf V}f_i^{(1)}\right)
+\frac{1}{2}\zeta^{(1)}
\frac{\partial}{\partial {\bf V}}\cdot \left({\bf V}f_i^{(0)}\right)=
-{\cal L}_if_i^{(1)}-{\cal M}_if_j^{(1)},
\end{equation}
where the operators ${\cal L}_i$ and ${\cal M}_i$ are defined by 
Eqs.\ (\ref{a2}) and (\ref{a3}), respectively. Since $f_i^{(1)}$ depends on 
time only through the temperature, Eq.\ (\ref{4.1}) implies that
$\partial_t f_i^{(1)}={\cal O}(a^2)$ and so the first term on the left 
hand side of Eq.\  (\ref{b1}) vanishes in the first order. Further, 
$\zeta^{(1)}=0$ by symmetry because $\nabla \cdot {\bf u}=0$ in the USF. 
Taking into acount the above properties, Eq.\ (\ref{b2}) reduces to 
\begin{equation}
\label{b3}
\left({\cal L}_i+\frac{1}{2}\zeta^{(0)}\frac{\partial}{\partial {\bf V}}
\cdot {\bf V}\right)f_i^{(1)}+{\cal M}_if_j^{(1)}=aV_y \frac{\partial}{\partial V_x} 
f_{i}^{(0)}.
\end{equation} 
This integral equation is identical to Eq.\ (\ref{a5}) when one 
particularizes the latter one to USF.
Therefore, the expression for the shear viscosity obtained from (\ref{b3}) is given by 
Eqs.\ (\ref{3.5})--(\ref{3.9}).

\begin{figure}
\caption{Plot of the ratio $\eta^*(a^*)/\eta^*_{\text{B}}$
as function of $a^*$ for $\alpha=0.9$ in
the case $m_1/m_2=4$, $n_1/n_2=1$, and $\sigma_1/
\sigma_2=3$ for three different values of the initial shear rate $a_0^*$:
$a_0^*=0.2, 0.3,$ and $0.4$. Here, $\eta^*_{\text{B}}$ refers to the Navier-
Stokes shear viscosity value given by the first Sonine approximation to
the Boltzmann equation.}
\label{fig1}
\end{figure}

\begin{figure}
\caption{Plot of the temperature ratio $T_1/T_2$ as a function of the
size ratio $\sigma_1/\sigma_2=(m_1/m_2)^{1/3}$
for $n_1/n_2=1$ and three different
values of the restitution coefficient $\alpha$: (a) $\alpha=0.9$ 
(circles), (b) $\alpha=0.8$ (squares), and (c) $\alpha=0.7$ (triangles).
The lines are the theoretical predictions and the symbols refer to the results
obtained from the DSMC method.
}
\label{fig2}
\end{figure}

\begin{figure}
\caption{Plot of the ratio $\eta^*(\alpha)/\eta^*(1)$ as a function of the
mass ratio $m_1/m_2$ for $\sigma_1/\sigma_2=n_1/n_2=1$ and three different
values of the restitution coefficient $\alpha$: (a) $\alpha=0.9$ 
(circles), (b) $\alpha=0.8$ (squares), and (c) $\alpha=0.7$ (triangles).
The lines are the theoretical predictions and the symbols refer to the results
obtained from the DSMC method.
}
\label{fig3}
\end{figure}

\begin{figure}
\caption{Plot of the ratio $\eta^*(\alpha)/\eta^*(1)$ as a function of the
size ratio $\sigma_1/\sigma_2$ for $m_1/m_2=4$, $n_1/n_2=1$
and three different
values of the restitution coefficient $\alpha$: (a) $\alpha=0.9$ 
(circles), (b) $\alpha=0.8$ (squares), and (c) $\alpha=0.7$ (triangles).
The lines are the theoretical predictions and the symbols refer to the results
obtained from the DSMC method.
}
\label{fig4}
\end{figure}

\begin{figure}
\caption{Plot of the ratio $\eta^*(\alpha)/\eta^*(1)$ as a function of the
concentration ratio $n_1/n_2$ for $m_1/m_2=4$, $\sigma_1/\sigma_2=1$
and three different
values of the restitution coefficient $\alpha$: (a) $\alpha=0.9$ 
(circles), (b) $\alpha=0.8$ (squares), and (c) $\alpha=0.7$ (triangles).
The lines are the theoretical predictions and the symbols refer to the results
obtained from the DSMC method.
}
\label{fig5}
\end{figure}

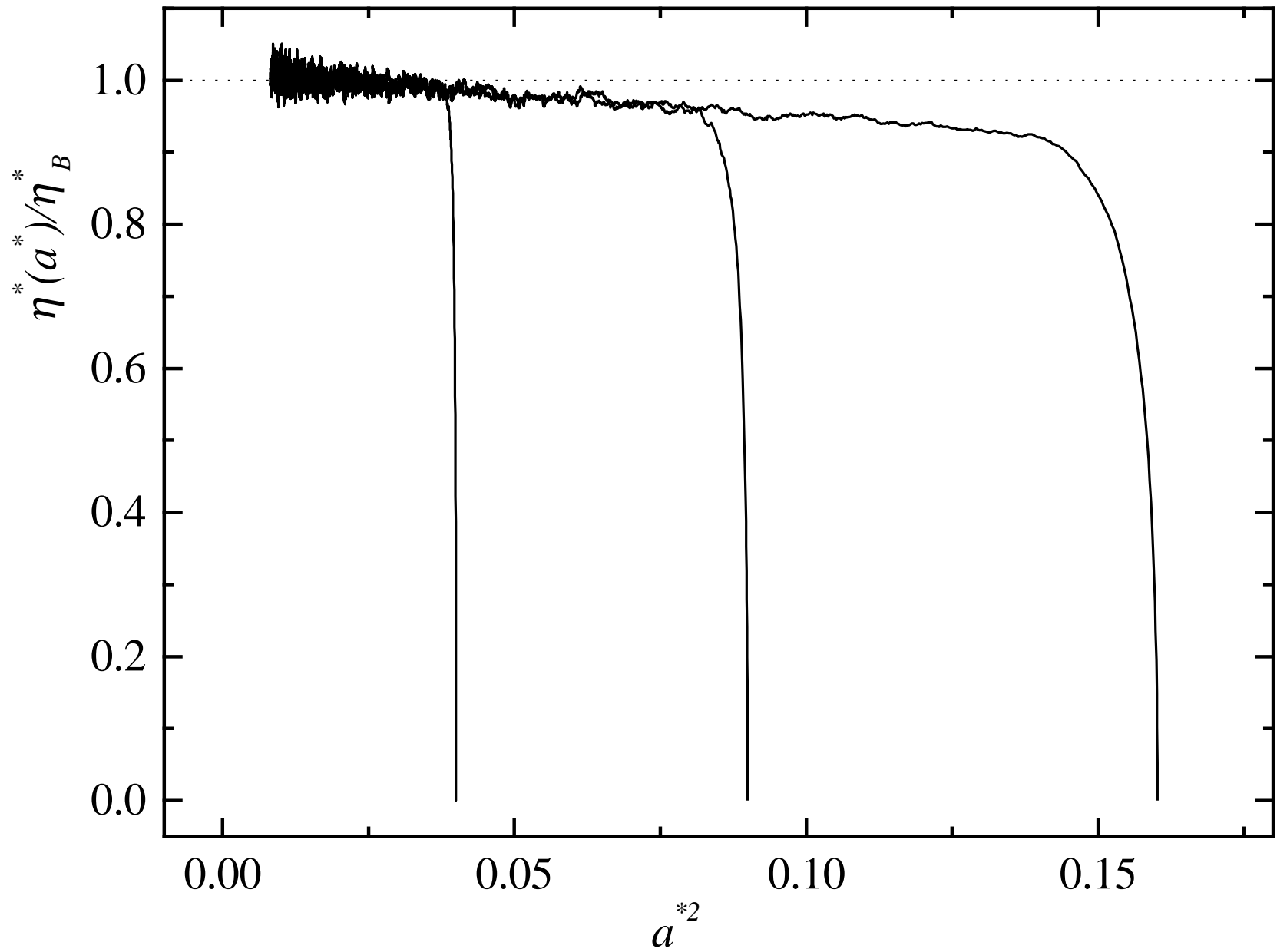
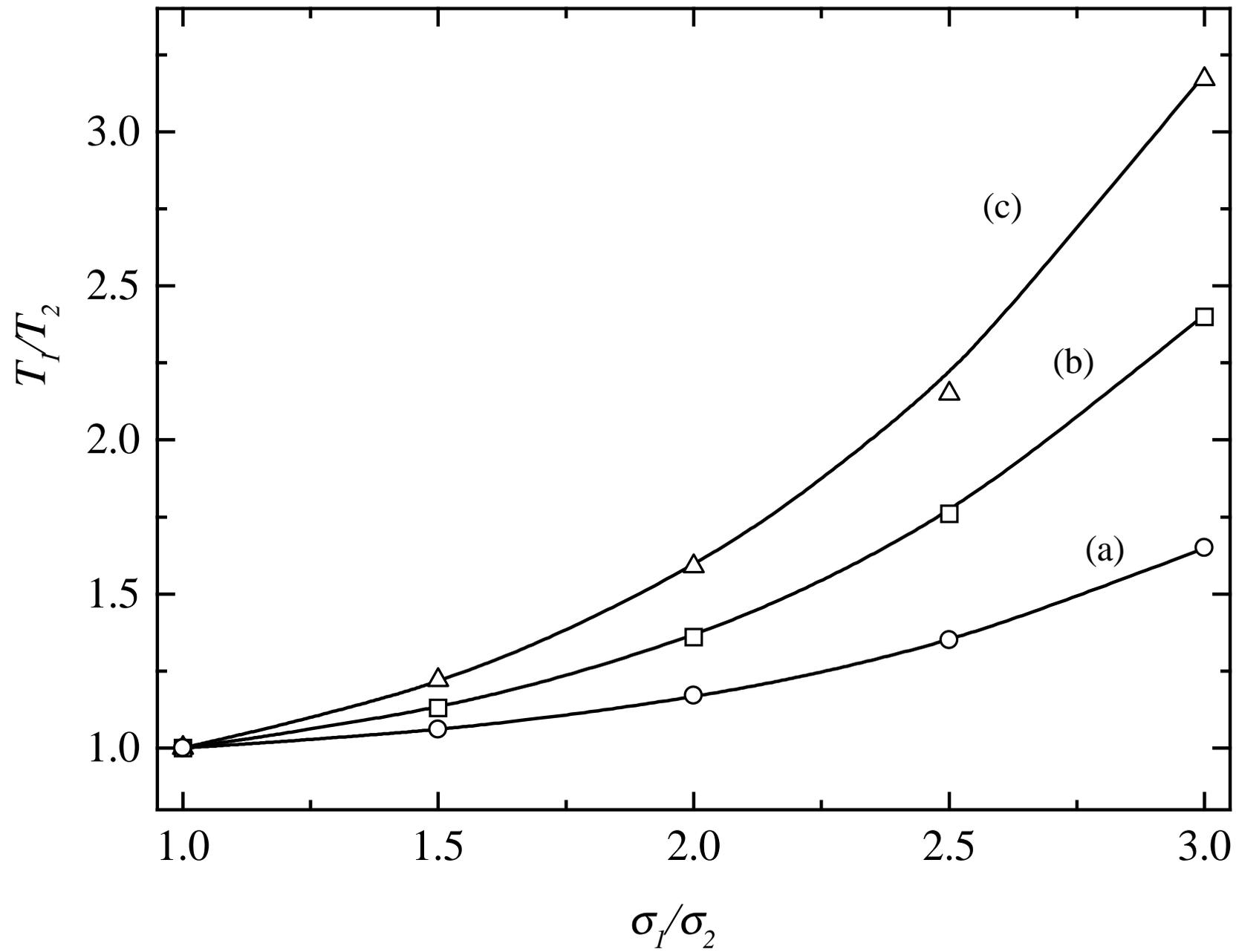
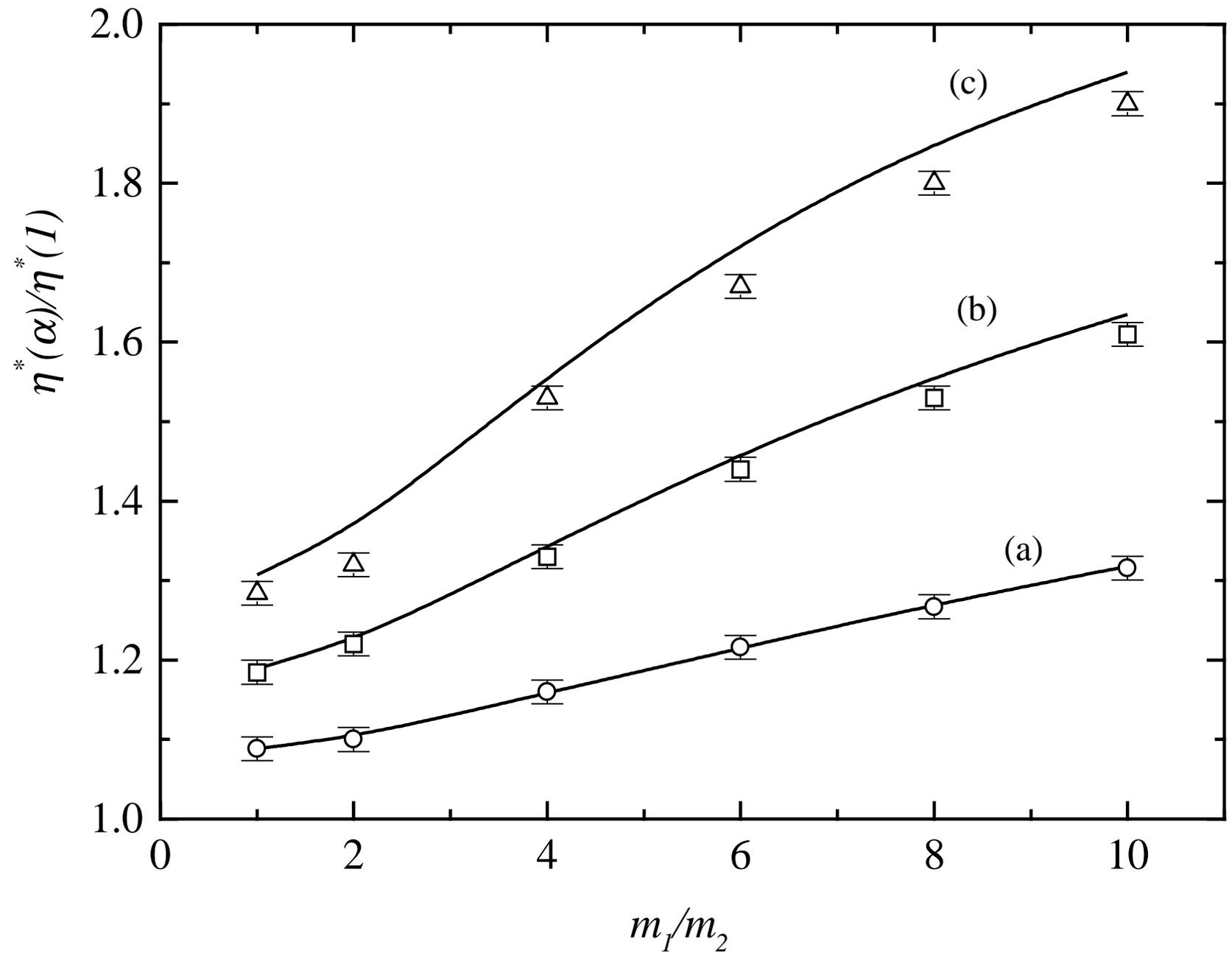
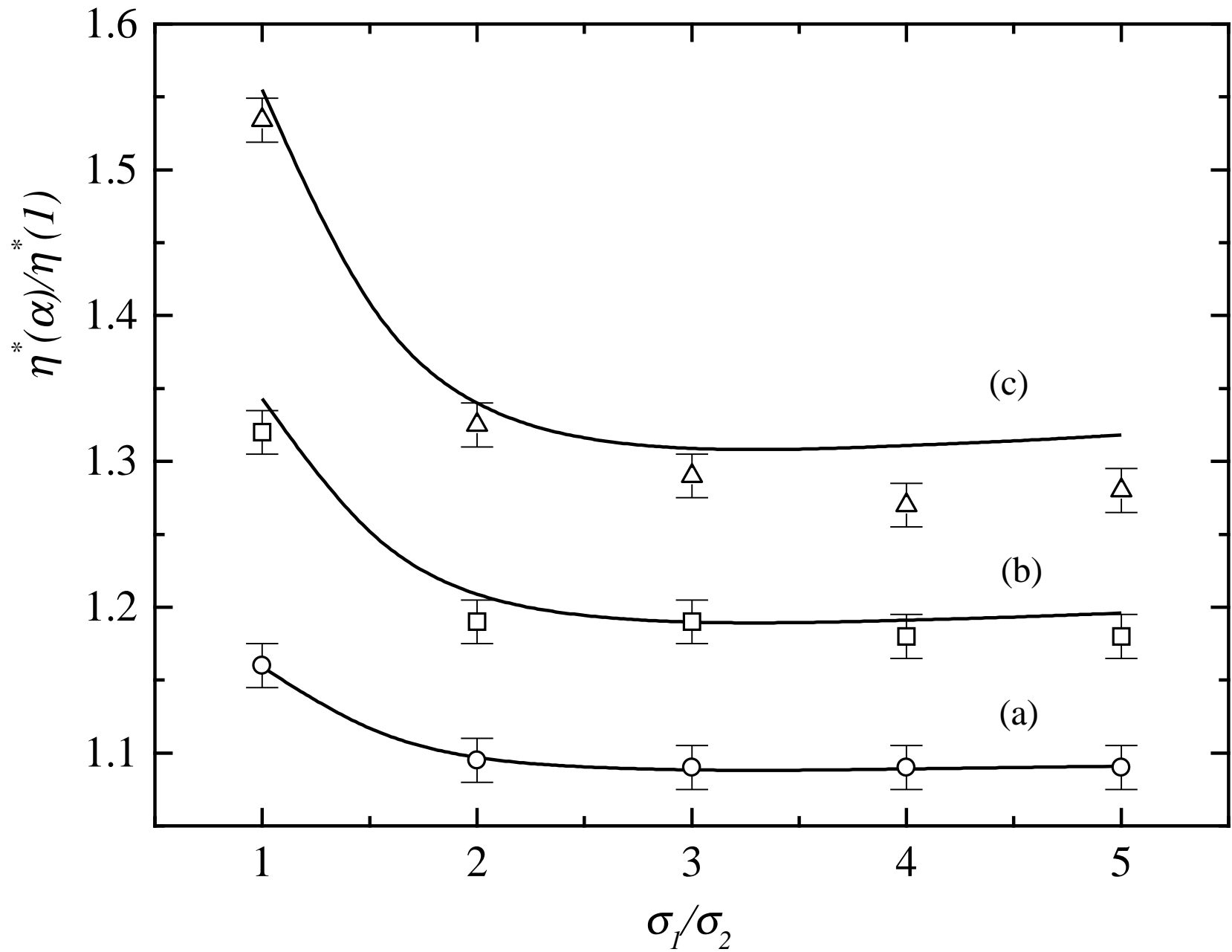
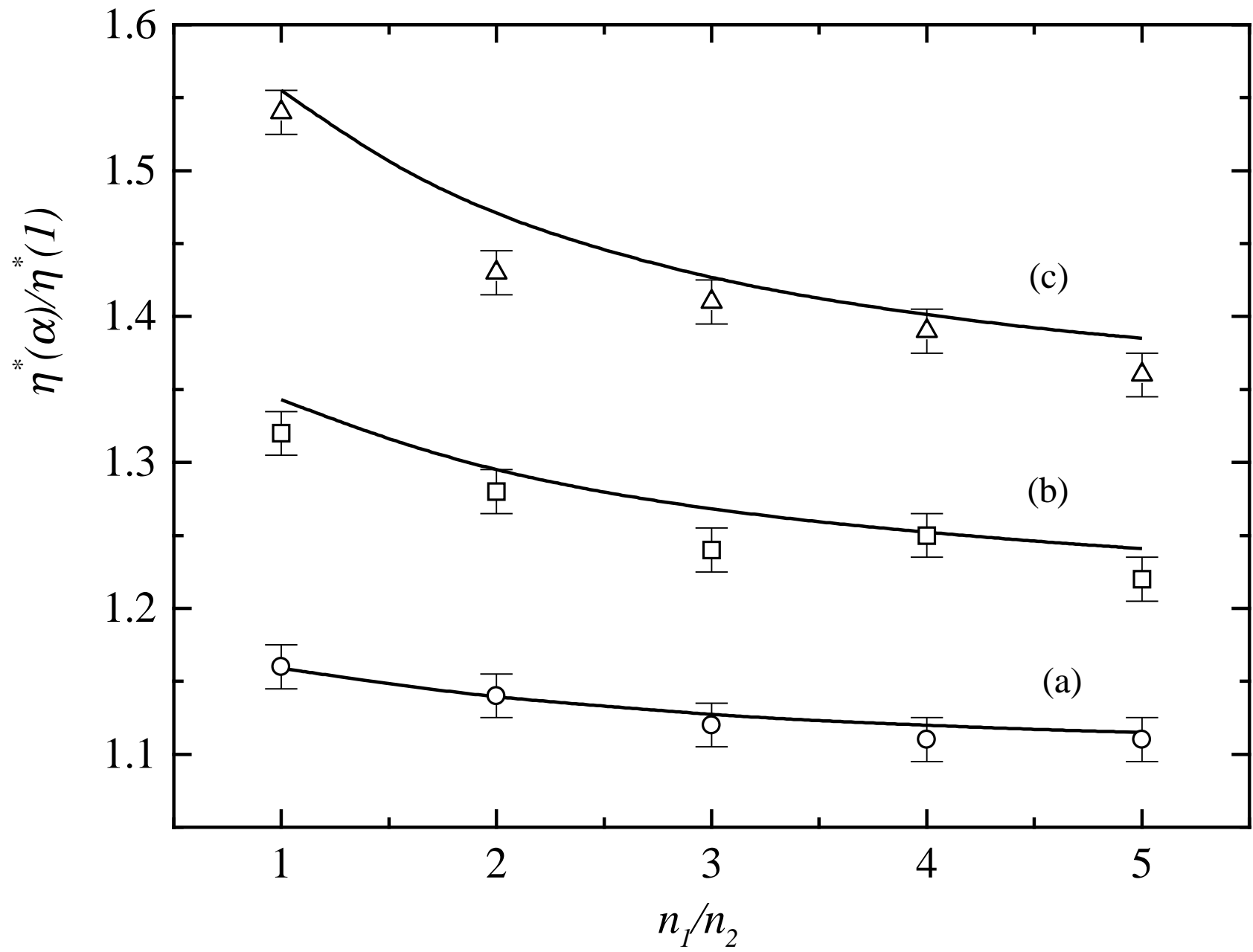
\end{document}